\documentclass[aps,prl,reprint,runinaddress,showpacs,showkeys,nofootinbib]{revtex4-1}

\usepackage[hyperfootnotes=false,colorlinks=true,citecolor=blue]{hyperref} 
\hypersetup{pdfauthor={Shalaev, Desnavi, Walasik, Litchinitser},pdftitle={Reconfigurable Topological Photonic Crystal}}
\usepackage{amsmath}
\usepackage{amsfonts}
\usepackage{textcomp}
\usepackage[]{graphicx}
\usepackage{epstopdf}
\usepackage{mathrsfs}
\usepackage{amsbsy}
\usepackage{cleveref}
\usepackage{units}
\usepackage{MnSymbol}
\usepackage{lipsum}
\usepackage{color}
\usepackage{easybmat}
\usepackage{natbib}

\usepackage[export]{adjustbox}
\usepackage{soul}

\crefformat{equation}{Eq.~(#2#1#3)}
\crefformat{figure}{Fig.~#2#1#3}

\Crefformat{equation}{Equation~(#2#1#3)}
\Crefformat{figure}{Figure~#2#1#3}

\begin{document}
	
	\title{Reconfigurable Topological Photonic Crystal}
	
	\author{Mikhail I. Shalaev$^*$}
	\affiliation{Department of Electrical Engineering, University at Buffalo, The State University of New York, Buffalo, New York 14260, USA}
	\author{Sameerah Desnavi$^*$}
	\affiliation{Department of Electrical Engineering, University at Buffalo, The State University of New York, Buffalo, New York 14260, USA}
	\author{Wiktor Walasik$^*$}
	\affiliation{Department of Electrical Engineering, University at Buffalo, The State University of New York, Buffalo, New York 14260, USA}
	\author{Natalia M. Litchinitser$^{\dagger}$}
	\affiliation{Department of Electrical Engineering, University at Buffalo, The State University of New York, Buffalo, New York 14260, USA}

	\graphicspath{ {./Figures/} }
	
	\date{\today}
	
	\begin{abstract}
		Topological insulators are materials that conduct on the surface and insulate in their interior due to non-trivial topological order. 
		The edge states on the interface between topological (non-trivial) and conventional (trivial) insulators are topologically protected from scattering due to structural defects and disorders. 
		Recently, it was shown that photonic crystals can serve as a platform for realizing a scatter-free propagation of light waves. 
		In conventional photonic crystals, imperfections, structural disorders, and surface roughness lead to significant losses. 
		The breakthrough in overcoming these problems is likely to come from the synergy of the topological photonic crystals and silicon-based photonics technology that enables high integration density, lossless propagation, and immunity to fabrication imperfections.
		For many applications, reconfigurability and capability to control the propagation of these non-trivial photonic edge states is essential. 
		One way to facilitate such dynamic control is to use liquid crystals, which allow to modify the refractive index with external electric field. 
		Here, we demonstrate dynamic control of topological edge states by modifying the refractive index of a liquid crystal background medium. 
		Background index is changed depending on the orientation of a liquid crystal, while preserving the topological order of the system. 
		This results in a change of the spectral position of the photonic bandgap and the topologically protected edge states.
		The proposed concept might be implemented using conventional semiconductor technology, and can be used for robust energy transport in integrated photonic devices, all-optical circuity, and optical communication systems.
	\end{abstract}

	\pacs{03.65.Vf, 42.70.Qs, 42.79.Kr}
	\keywords{Topological phases (quantum mechanics), Photonic bandgap materials, Liquid crystals in optical devices}

	\maketitle
	
	\section*{Introduction}
	
	Topological insulators (TIs) build a class of materials that act as insulators in their interior and conduct on the surface, while having a non-trivial topological order~\cite{kane05a,katmis16,fereira13,moore10,kane05}. 
	The insulating properties result from the absence of conducting bulk states in a certain energy range, known as the bandgap. 
	The interface between materials with different topological order supports strongly confined topologically protected edge states. 
	For these states, the energy transport is robust against structural disorders and imperfections that do not change the system's topology. 
	Until now, many theoretical and experimental demonstrations of TIs have been reported for fermionic (electronic) systems, but most of them work at low temperatures and require strong external magnetic fields which impedes their practical applications~\cite{kane05a,kane05,bernevig06,konig07,hsieh08,zhang09}. 
	{\color{black}Alternatively, system preserving the time-reversal symmetry that support spin- and valley-Hall effects have been implemented~\cite{doi:10.1143/JPSJ.77.031007,valleyh}.}\let\thefootnote\relax\footnote{$^*$These authors made equal contributions.}\let\thefootnote\relax\footnote{$^{\dagger}$Corresponding author: natashal@buffalo.edu}
	
	Recent studies have shown the existence of one-way protected topological edge states in bosonic (photonic) systems. With the use of time-reversal symmetry breaking, an analogue of the quantum-Hall effect was achieved~\cite{haldane08,raghu08,wang08,wang09,poo11}.
	Later, analogues for spin-Hall and valley-Hall effects that do not require breaking of the time-reversal symmetry were realized using photonic TIs~(PTIs)~\cite{ma15,ma16,barik16,wu15}.
	A considerable interest has been shown in manipulating photons by the use of an artificial gauge field, which acts as an effective magnetic field
	for photons~\cite{fang12,umucaliar}. 
	Several approaches to engineer synthetic gauge potential emulating an effective magnetic field have been realized~\cite{fang12,umucaliar,hafezi11,hafezi13} by dynamic modulation of the system parameters. Examples  include  temporal~\cite{fang11,fang12,fang13,reiskarimian} and spatial modulation using an array of helical waveguides that imitate a breaking of the time-reversal symmetry by breaking the mirror symmetry along the propagation direction~\cite{rechtsman13}. 
	Some of the other proposed realizations include use of meta-crystals~\cite{khanikaev13,chen2014,jacobs15,cheng16,he16,slobozhanyuk16}. 
	Although these approaches demonstrate the possibility to realize PTIs, most of the designs either operate in the microwave regime or are bulky.
	Implementation of TIs in photonic systems can pave the way for robust light propagation unhindered by the influence of {\color{black}back-scattering} losses.

	The breakthrough is likely to come from the synergy of the PTI concept and silicon-based photonics technology that enables high integration density, reconfigurability, and immunity to fabrication imperfections. 
	In particular, silicon-based photonic crystals (PCs) offer a promising solution to integration of the fields of silicon photonics and topological photonics~\cite{ma16,barik16,wu15}.
	Indeed, PCs enable implementation of topological effects.
	Nowadays, the majority of proposed PTIs operate in a fixed wavelength range and their mode of operation cannot be dynamically reconfigured at a high speed. 
	Here, we propose a reconfigurable PTI structure based on PC design to realize the photonic analogue of the spin-Hall effect. 
	The tunability of transmission properties for the system is facilitated by the liquid crystal~(LC) environment surrounding the PC. 
	This structure offers compatibility with CMOS integrated systems, allows for switching at MHz frequencies, and can be designed to operate at telecommunication wavelengths. 
	
	\onecolumngrid
	\begin{center}
		\begin{figure}[!t]
			\includegraphics[width = \textwidth]{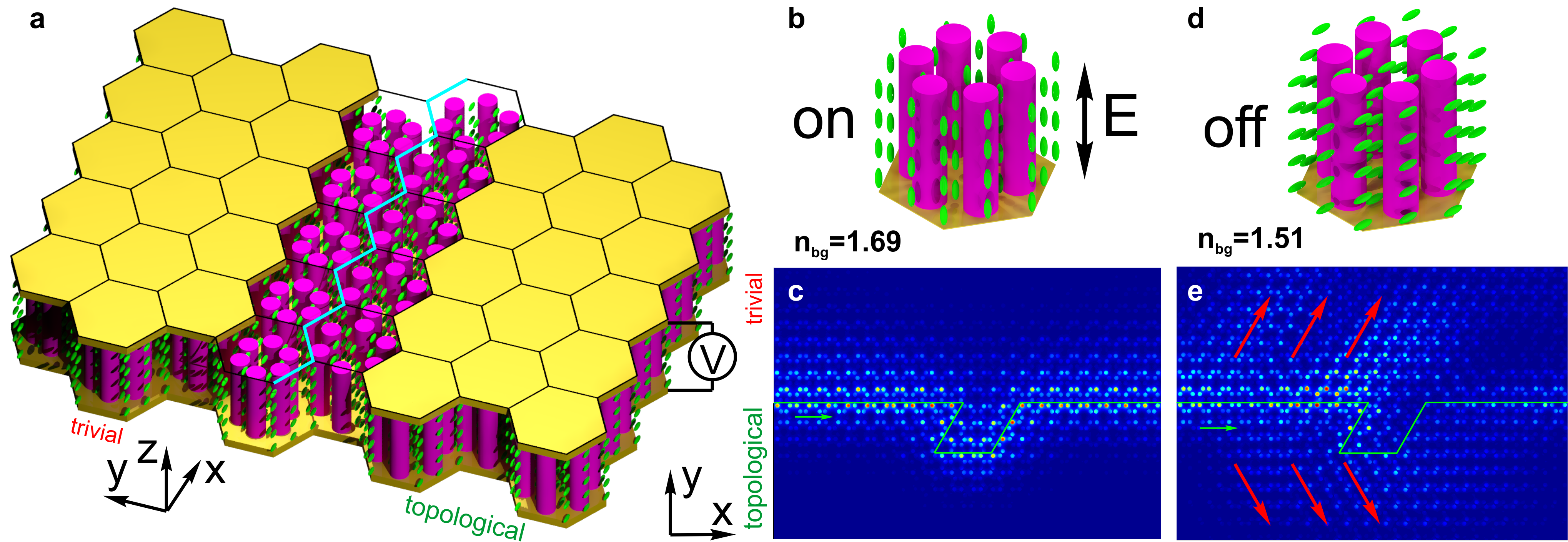}
			\caption{
				\textbf{{\color{black}Schematic view of the structure and working principle of the reconfigurable photonic-crystal-based topological insulator.}} 
				(a) The structure consists of silicon pillars (magenta) surrounded by a liquid crystal molecules (green) enclosed between {\color{black}conducting} electrodes (yellow). 
				It can be switched by applying voltage to the electrodes, such that when the voltage is applied, LC molecules orient along the pillars, resulting in the background refractive index $n_{\textrm{bg}}=1.69$ (b) for transverse-magnetic polarization considered here (the corresponding electric field polarized along the pillars is indicated by~$\updownarrow$E). 
				(c) The energy density distribution for the case when switch is ON: light is guided along rhombus-shaped path shown in green. 
				(d) When there is no voltage applied, background refractive index, $n_{\textrm{bg}}=1.51$, corresponding to the OFF-state, (e) Power penetrates to the crystal interior (shown in red) resulting in low transmittance along the interface path.}
			\label{fig:fig1}
		\end{figure} 
	\end{center}
	\twocolumngrid
		
	\section*{Results}
	
	Photonic crystals offer an excellent platform to control the flow of light by virtue of the periodicity of the dielectric constants of its constitutive materials~\cite{jannopoulos08}.  
	The backbone of the proposed structure is the PC built of silicon pillars immersed in LC environment and enclosed between {\color{black}conducting} electrodes; as shown schematically in~\cref{fig:fig1}(a). 
	{\color{black}The design of the PC providing topological protection is based on the work of Wu and Hu~\cite{wu15}.}
	The structure consists of two regions: one with trivial order, and another with topological. 
	At the interface between crystals with different topological properties (shown in cyan), edge states are supported.
	Each region is built of a triangular lattice containing six pillars per unit cell that build a meta-molecule. Depending on the spacing between the pillars, the structure features a topological or a trivial order.
	Topological edge states for this type of PC emerge due to the optical analogue of the spin-Hall effect, and were analyzed in Ref.~\cite{wu15} where the PC was surrounded by air.

	The PC in our design is immersed in a nematic LC environment, which offers a possibility of refractive index tuning with unprecedented amplitude, reaching 10\%~at MHz switching speed~\cite{doi:10.1080/15421406.2011.569456,PhysRevLett.111.107802,Mohammadimasoudi:14}. This tuning is enabled by an external electric field supplied by the electrodes bounding the PC from top and bottom~\cite{khoo93,weirich10}.
	Here, we restrict ourself to considering transverse-magnetic-polarized waves at telecommunications wavelength 1.55~$\mu$m and assume the use of the $E7$ nematic LC~\cite{li05b}.
	When voltage is applied, corresponding to ON-state, the LC molecules align along the silicon pillars, resulting in a background refractive index $n_{\textrm{bg}} = n_e = 1.69$ (extraordinary refractive index) [\cref{fig:fig1}(b)].
	In this case, light is efficiently guided along a rhombus-shaped path {\color{black}in an edge state located in the bulk bandgap,} as shown in~\cref{fig:fig1}(c).
	If there is no voltage applied, corresponding to the OFF-state, the  LC molecules orient perpendicular to the pillars due to the anchoring forces at the electrodes, and light experiences the background refractive index $n_{\textrm{bg}} = n_o = 1.51$ (ordinary refractive index), as shown~\cref{fig:fig1}(d). 
	{\color{black} The change of the background index does not modify the topological properties of the structure, but shifts the location of the bandgap, and bulk states are present at the wavelength of interest, enabling light scattering into the bulk of the PC.}
	For this case, the structure does not support light propagation and energy penetrates into the PC interior resulting in low transmittance along the interface, as shown in~\cref{fig:fig1}(e).

	\subsection*{Tunable topological edge states}
	
	Topologically protected edge states offer unprecedented possibilities for designing robust guided-wave photonic structures and components, due to their feasibility for supporting light propagation along arbitrary-shaped interfaces between trivial and topological regions. 
	In this paper, we consider a structure exhibiting an optical equivalent of the spin-Hall effect. For this case, there 
	is always a pair of essentially decoupled states that propagate in opposite directions and have different spins. 
	In	sharp contrast to edge states in standard (trivial) PCs, the pair of topological states do not couple to each other even in presence of disorders and sharp turns along the propagation path.

	Let us consider scattering of light in standard PCs by obstacles on the light's path, such as rapid turns, crystal imperfections, or defects. 
	When light propagating in the forward direction impinges on an obstacle, the wavevectors matching backward-propagating state may be introduced. 
	For standard PC, the field distributions of forward- and backward-propagating states possess inversion symmetry along the propagation path, resulting in significant scattering of light from the forward state to the backward one~\cite{ma15,barik16,wu15,xu16}, degrading the PC performance and resulting in scattering losses. 
	On the contrary, for topologically protected states, the field shows vortex-like distribution for forward- and backward-propagating states (opposite spins), breaking the inversion symmetry. 
	In this case, the field distributions of opposite spins do not overlap, and therefore the states do not scatter one to another, resulting in the suppression of backward scattering for the photonic analog of spin-Hall effect.

\begin{figure}[!t]
		\includegraphics[width = \columnwidth, clip = true, trim = {0 0 0 0}]{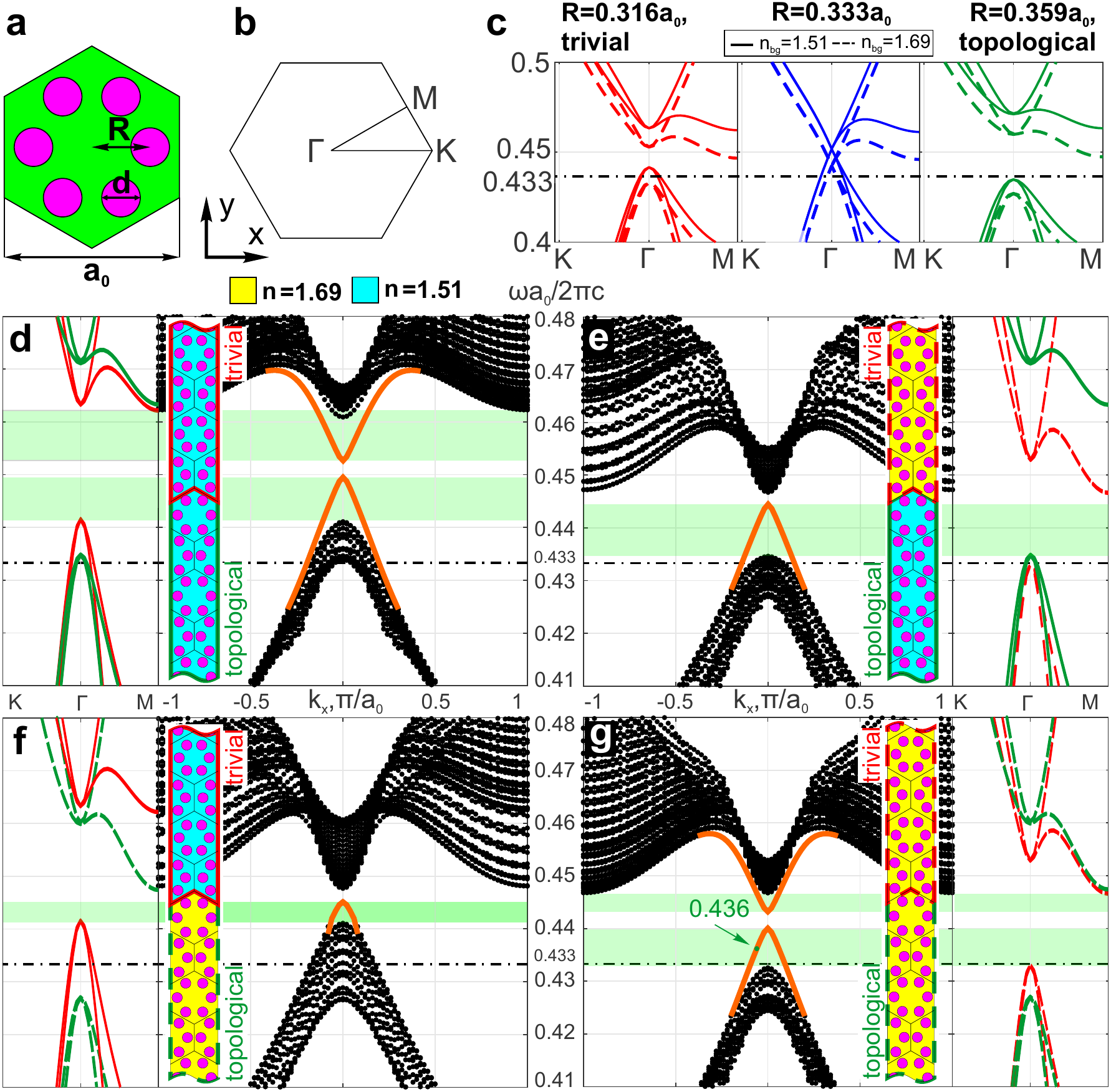}
		\caption{
			\textbf{Dispersion relations for the trivial and topological photonic crystals and the edge state diagrams at the interfaces with different topological orders.}
			(c) Band diagrams for trivial and topological PCs with hexagonal unit cell (a) and different background refractive indices. The first Brillouin zone of the triangular lattice is shown in panel (b). 
			Increase of the distance between the meta-molecule center and the pillars, $R$, leads to stronger coupling between neighboring meta-molecules. This results in the opening of the bandgap with a non-trivial topological order.
			In contrast, for smaller $R$, the bandgap opens while maintaining trivial topological order.  
			(d)--(g) Dispersion relations for ribbon-like photonic crystals formed by two regions with trivial and non-trivial topology (single periods along the $x$-direction are shown in the insets).
			The mechanism of formation of a bulk bandgap for this case is illustrated in the outside panels by combining the band structure diagrams for bulk topological and trivial crystals with triangular lattice (c) and corresponding refractive indices. 
			Green shading shows the frequency range where topologically protected propagation is supported: in this case an edge state (shown in orange) exists while no bulk states are present.
			The field distributions for spin-down states with normalized frequency $\omega a_0/(2\pi c)=0.433$ propagating along a rhombus-like interface between trivial and topological regions are presented in the corresponding plots in~\cref{fig:rhombus}
		}
		\label{fig:fig2}
	\end{figure}

\begin{figure}[!t]
		\includegraphics[width = \columnwidth, clip = true, trim = {0 20 0 50}]{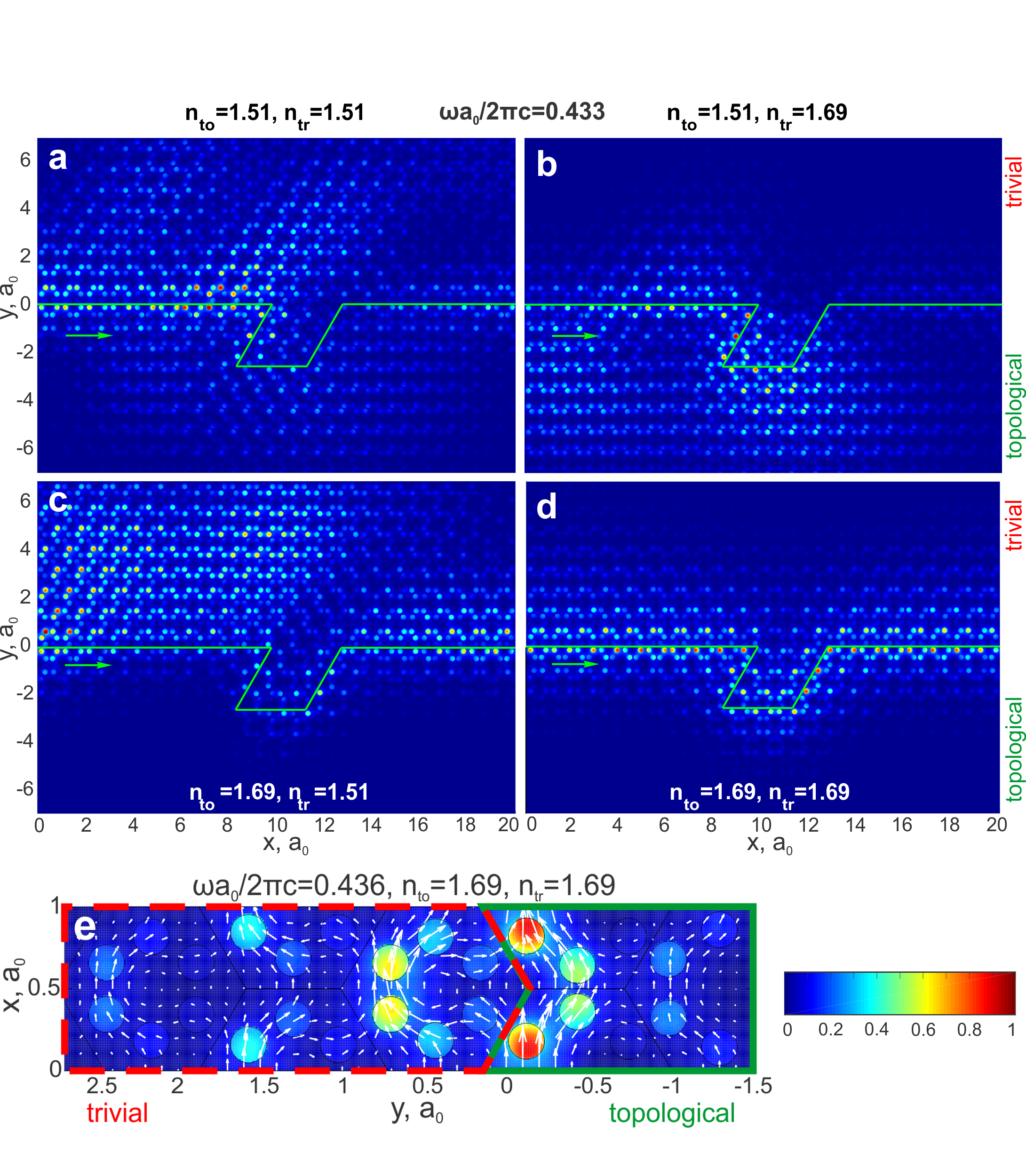}
		\caption{
			\textbf{Light propagation along the interface between trivial and topological photonic crystals for different values of background refractive indices.}
			(a)--(d) Energy density distributions of the spin-down states at normalized frequency $\omega a_0/(2\pi c) = 0.433$, indicated by the dash-dot line in~\cref{fig:fig2}, for the four different configurations of background index values described in~\cref{fig:fig2}. 
			The color maps show that topologically protected propagation is supported only for the case (d) where there is no bulk state allowed for both topological and trivial regions at the considered frequency, as shown in the right panel of~\cref{fig:fig2}(g). 
			For the cases shown in (b) and (c), light penetrates inside the topological and trivial parts, respectively, due to the presence of bulk states for these regions, as shown in the right and left panels in Figs.~\ref{fig:fig2}(e) and (f), respectively. 
			Bulk states are allowed in non-trivial and trivial regions for the case (a) resulting in light penetration to both regions [see~\cref{fig:fig2}(d)].
			(e) Energy density (color map) and the Poynting vector (white arrows) for a spin-down mode in the structure with $n_{\textrm{tr}} = n_{\textrm{to}} = 1.69$ at the normalized frequency $\omega a_0/(2\pi c) = 0.436$, corresponding to the mode indicated by the green point in~\cref{fig:fig2}(g). 
		}
		\label{fig:rhombus}
	\end{figure}

    Here, we define the conditions required for achieving such scatter-free propagation.
	Firstly, at the desired frequency, a non-trivial edge state should exist. To confirm the presence of an edge state, we considered a ribbon-shaped PCs shown schematically in the insets in~\cref{fig:fig2}. 
	The band structure of this system reveals the presence of the bulk and edge states. 
	However, the existence of an edge state is not sufficient to have loss-free propagation. 
	Indeed, if besides the non-trivial edge state there is at least one bulk state at the frequency of interest, any obstacle on the light propagation path will cause undesired scattering of light into the PC interior, resulting in losses of energy, as shown in Figs.~\ref{fig:fig2} and~\ref{fig:rhombus}. 
	Hence, the second necessary condition is the absence of any bulk states at the frequency where topologically-protected propagation is desired. 
	
	
	{\color{black}Light can be confined along the $z$-direction using two alternative mechanisms. The use of metal electrodes allows for strong light confinement inside the PC slab, however, this approach introduces an additional source of loss in the system at optical frequencies. Alternatively, the electrodes can be located at a distance from the PC slab. The space between electrodes and the slab can be filled with the LC which has much lower index than the PC itself. In this case, the light confinement stems from the total internal reflection between the high-index silicon slab and the low-index cladding. In both cases, these 3-dimensional~(3D) systems can be well approximated using 2D analysis.}

	Figures~\ref{fig:fig2}(d) and~(g) present the band diagrams for the interface between the trivial and topological regions for uniform background refractive indices $n_{\textrm{tr}} = n_{\textrm{to}} = 1.51$ and $n_{\textrm{tr}} = n_{\textrm{to}} =1.69$, respectively. 
	In both cases, two edge states---one corresponding to a pseudo-spin-up (denoted by a minus sign and simply referred to as spin-up in the following text) and one corresponding to a pseudo-spin-down (denoted by a plus sign and referred to as spin-down)---are present in the bandgap separating the bulk bands{\color{black}, showing that the change of the background refractive index does not change the topological properties of the system}. 
	{\color{black}There is a small frequency range where neither the edge states nor the bulk states exist. 
	This global gap is a result of the avoided crossing of the edge states caused by their interaction due to the broken $C_6$ symmetry at the conducting interface~\cite{wu15,barik16,xu16}.}
	The position and size of the bandgap is affected by variations of the background refractive index. 
	For the case of background index of 1.51, shown in~\cref{fig:fig2}(c), the bandgap spans the normalized frequency range $\omega a_0/(2\pi c)\in [0.441,  0.462]$, whereas for the background index of 1.69 shown in~\cref{fig:fig2}(g), the bandgap extends from 0.433 to 0.447. 
	The shrinking and red-shifting of the bandgap observed here is consistent with the results presented in Ref.~\cite{guryev06}. 
	The effective working wavelength in the medium stays the same, but as a result of an increase in the effective index, the corresponding frequency (free space wavelength) becomes lower (higher)~\cite{jannopoulos08}.

	We choose to analyze the behavior of the structure at a normalized frequency of 0.433 that is located outside of the bandgap for $n_{\textrm{bg}}=1.51$, and inside of the bandgap for $n_{\textrm{bg}}=1.69$. Figures~\ref{fig:rhombus}(a) and~(d) show the propagation of light along an interface with a rhombus-shaped defect at $\omega a_0/(2\pi c)=0.433$ for background refractive indices of 1.51 and 1.69, respectively. 
	When the background index has the value of 1.51, the edge state does not exist and the light couples to the bulk modes of both the trivial region located on the top of the structure, as well as the topological region on the bottom. 
	This behavior can be understood by looking at the band structures for infinitely periodic triangular PCs shown in the left and right panels in Figs.~\ref{fig:fig2}(d) and~(g), respectively. 
	For $n_{\textrm{bg}}=1.51$ (see solid curves), the normalized frequency 0.433 is located below the bandgap for both trivial (red curve) and topological (green curve) PC geometry. 
	On the contrary, this frequency is located inside the bandgap of both trivial (red curve) and topological (green curve) structures for $n_{\textrm{bg}}=1.69$ [dashed curves in the right panel of~\cref{fig:fig2}(g)]. 
	Therefore, as seen in~\cref{fig:rhombus}(d), the light propagates along the rhombus-like shaped interface between trivial and topological material without scattering to the bulk. 
	The electric field and the Poynting vector distributions in the vicinity of the waveguiding interface for the spin-down eigen-mode of the system with uniform background index $n_{\textrm{bg}}=1.69$ at the normalized frequency $\omega a_0/(2\pi c)=0.436$ is shown in~\cref{fig:rhombus}(e), indicating a strong light localization at the interface between the trivial and topological PC configurations. 
	Moreover, the Poynting vector shows the vortex-like character of energy propagation along the interface associated with the spin-Hall-effect nature of the TI.

	\begin{figure*}[t]
		\includegraphics[width = 0.8\textwidth, clip = true, trim = {0 0 0 0}]{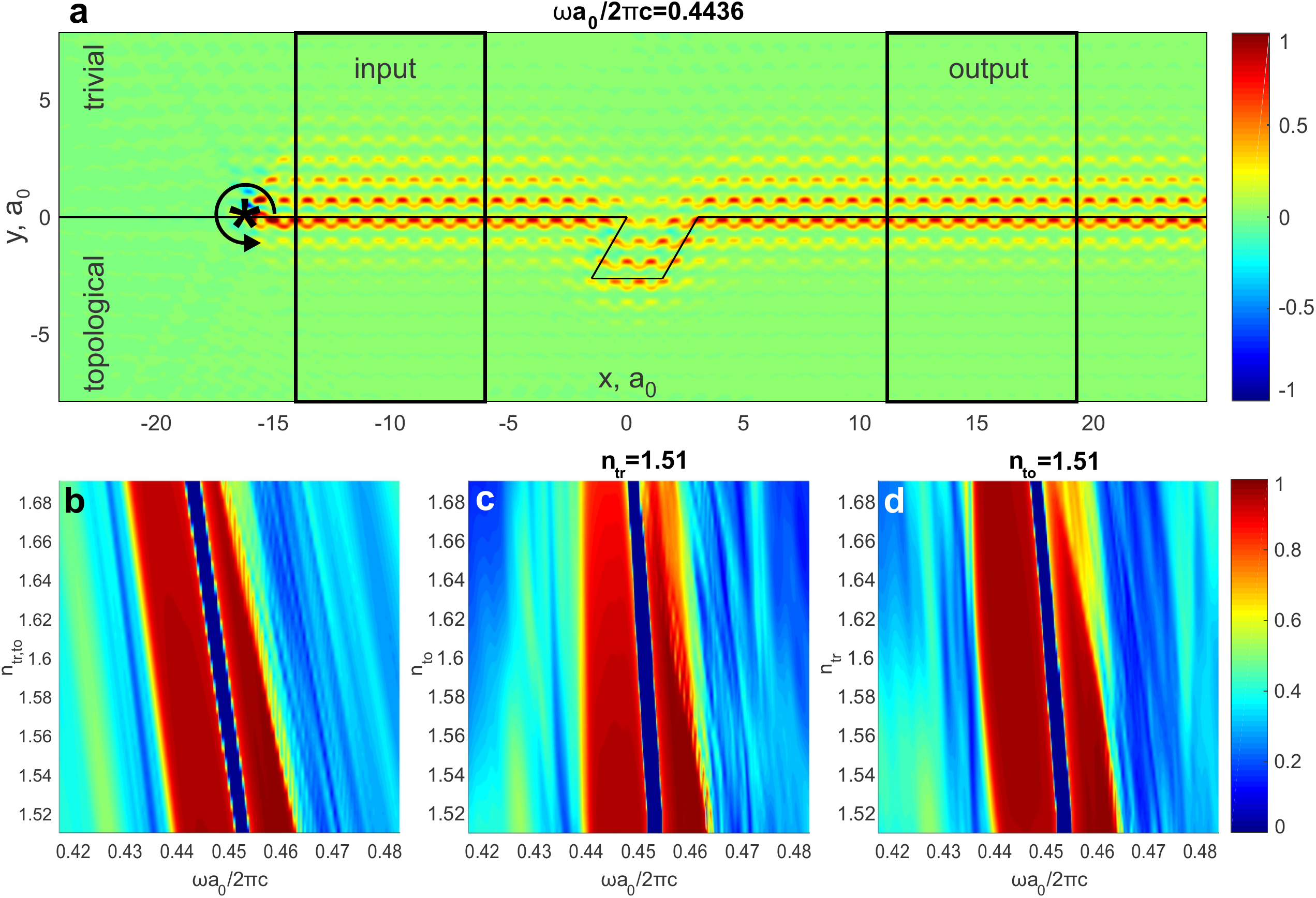}
		\caption{
			\textbf{Transmission characteristics of the reconfigurable topological insulator.} 
			(a) Normalized Poynting vector distribution along the $x$-axis, $P_x$, at the normalized frequency $\omega a_0/(2\pi c) = 0.4436$ showing light propagation along a rhombus-shaped interface between the trivial and topological photonic crystals. The light is excited with the magnetic dipole source $H_{+} = H_x + iH_y$ (indicated with a star). 
			(b)--(d) Transmission spectrum dependent on background refractive index for three different cases: (b) background refractive index varied in both the topological and trivial regions simultaneously; (c) background index is varied in the topological part and is fixed in the trivial one $n_{\textrm{tr}}= 1.51$; (d) background index is fixed in the topological region $n_{\textrm{to}}= 1.51$ and varies in the trivial part. 
			For all cases, with the increase of the background refractive index, the index contrast in the structure decreases, leading to the narrower transmission bandwidth.
			Furthermore, for higher background indices, the average refractive index increases, resulting in the red-shift of the spectral location of the guided region.}
		\label{fig:trans}
	\end{figure*}

	We have also analyzed the behavior of mixed configurations where the background refractive indices of the trivial and topological regions have different values. 
	This can be achieved by independent control of the voltages on the electrodes sandwiching the trivial and topological regions. 
	To this end, the top electrodes should be separated by a thin insulating layer. 
	The band diagrams for these systems are presented in Figs.~\ref{fig:fig2}(e),~(f) and the corresponding light propagations through the rhombus-like defect is shown in Figs.~\ref{fig:rhombus}(b),~(c). 
	The band diagrams presented in~\cref{fig:fig2} show that the normalized frequency of interest, $\omega a_0/(2\pi c) = 0.433$, is located below the bandgap both for the configuration with $n_{\textrm{to}}=1.51$, $n_{\textrm{tr}}=1.69$  and the inverse configuration with  $n_{\textrm{to}}=1.69$, $n_{\textrm{tr}}=1.51$. 
	Therefore, in both cases, the light scatters into the bulk, and it penetrates the region with the lower background refractive index, as can be seen in Figs.~\ref{fig:rhombus}(b) and~(c). 
	This behavior again can be explained using the right and left panels in Figs.~\ref{fig:fig2}(e) and~(f), respectively, by analyzing whether the studied frequency lays inside or outside of the bandgap for the bulk trivial and topological PCs.  
	The examples described above show that the transmission of light in the edge states at the interface between the trivial and topological PCs can have drastically different character depending on the background refractive index. 
	Therefore, the control scheme over the background index presented above enables us to modify the transmission properties of the system.
	
	\subsection*{Transmission properties of the reconfigurable topological structure}
	
	Let us consider light propagation along an interface between the trivial and topological regions with a 60\textdegree\ rhombus-shaped path.
	Transmission calculation through the rhombus allowed us to identify spectral positions of the topologically protected guided modes for different configurations of background refractive indices in trivial ($n_{\textrm{tr}}$) and topological regions ($n_{\textrm{to}}$).
	The transmission spectrum is calculated according to the following expression:
	\begin{equation}
	T(\omega) = \frac{\int_{\textrm{out}} P_x(\omega)\textrm{d}x\textrm{d}y }{\int_{\textrm{in}} P_x(\omega)\textrm{d}x\textrm{d}y }, 
	\label{eqn:trans}
	\end{equation}
	where $P_x$ denotes the $x$-component of the Poynting vector and the surface integrations are performed over the input and output regions shown in~\cref{fig:trans}(a). From the typical distribution of $P_x$ when the transmission is close to unity, as depicted in~\cref{fig:trans}(a), we observe that the energy is well localized near the edge.
	
	Here, we consider the dependence of the transmission spectrum on the background refractive index variation for three cases: (i) background index is simultaneously changed in the topological and trivial regions [\cref{fig:trans}(b)]; (ii) background index is modified in the topological region and fixed in the trivial region, $n_{\textrm{tr}}= 1.51$  [\cref{fig:trans}(c)]; (iii) background index is kept constant in the  topological region, $n_{\textrm{to}}= 1.51$, and varied in the trivial region [\cref{fig:trans}(d)]. 
	Spectral position of topologically protected modes is defined by the top and bottom frequencies of the bulk bandgaps in both topological and trivial regions. These frequencies are dependent on the background refractive index, as shown in~\cref{fig:fig2}.

	Figures~\ref{fig:trans}(b)--(d) show that when the topologically protected guiding conditions (existence of edge state and lack of bulk states at given frequency) are satisfied, the transmission is high and close to unity. 
		All three of the plots show the red-shift of the guided region with an increase in refractive index, due to the higher average index values. The width of the guided region is decreased for higher background indices as a result of reduced refractive index contrast between the silicon pillars and the background. 
	
	Choosing an appropriate operation frequency allows for switching between high and low transmission modes. For instance, propagation at the normalized frequency $\omega a_0/(2\pi c)=0.436$ results in low transmission for $n_{\textrm{to}}=n_{\textrm{tr}}=1.51$, and topologically protected quasi-unitary transmission for $n_{\textrm{to}}=n_{\textrm{tr}}=1.69$. Choosing $a_0=675$~nm results in operational wavelength of 1550~nm. Alternatively, one could operate at $\omega a_0/(2\pi c)=0.457$ where the high and low transmission modes are reversed.

	\section*{Conclusion}
	
	In this paper, we have proposed a dynamically tunable topological photonic crystal enabled by the photonic analog of the spin-Hall effect. 
	The structure supports edge states at the interface between the trivial and topological parts of the crystal. These edge states are topologically protected and robust against structural disorders and imperfections. Their propagation is supported along arbitrarily shaped paths and around defects. 
	The reconfigurability is facilitated by immersing the photonic crystal into a nematic liquid crystal background.
	With the help of an external field applied to the liquid crystal, its molecules can be reoriented, causing variation in background refractive index and shifting the spectral position of edge states.
	We have shown that with rise of background permittivity, edge states exhibit red-shift due to rise in average refractive index of the crystal. 
	The transmission characteristics through the structure can be dynamically tuned by modifying the spectral position of the non-trivial bandgap.
	Moreover, the topologically protected bandwidth decreases with an increase of the background refractive index because of the reduction in the index contrast between the background and high-index material.
	We have defined the conditions that are necessary for supporting topologically protected propagation to be: the presence of non-trivial edge states, along with an absence of bulk state(s) at desired guided frequencies.
	When these conditions are satisfied, the structure supports topologically protected modes with transmittance close to 100\%. 
	Shifting the bandgap position results in scattering of light into the crystal interior, and a decrease in the transmittance through the structure.
	The reconfigurable photonic topological insulator proposed here is silicon based, and supports operation at telecommunication frequencies, making it attractive for practical applications. 
    {\color{black}An alternative mechanism for transmission control could be achieved by dynamical switching between trivial and topological states of the structure. This concept is outside of the scope of this paper and requires further investigation.}

	\section*{Methods}
	
	\subsection*{Band structure calculations}
	
	The band diagrams for bulk PCs with different background refractive indices shown in~\cref{fig:fig2}(c) were obtained using a plane-wave-expansion method. The band diagrams for the edge states shown in Figs.~\ref{fig:fig2}(d)--(g) and the energy distributions shown in~\cref{fig:rhombus} were calculated using COMSOL Multiphysics.  In order to compute the band diagrams for the edge states we have analyzed a ribbon-shaped PC infinitely periodic along the $x$-direction with a finite size of 30 unit cells of both trivial and topological regions along the $y$-direction.
	
	\subsection*{Transmission calculations}
	
	For transmission calculations, we used commercially available Lumerical FDTD Solutions software. The time domain calculations were carried in the simulation domain shown in~\cref{fig:trans}(a), and the spectral response was obtained by Fourier transform method. The simulation domain is $50 \times 20$ unit cells large, and is surrounded by perfectly-matched layers (PMLs). The size of the rhombus-shape path modification is $3 \times 3$ unit cells.
	
	The system was excited with a spin-down (right circularly polarized light rotating counter-clockwise) dipole point source $H_{+} = H_x + iH_y$ placed near the interface between trivial and topological parts of the crystal, matching well the profile of the mode propagating in the positive direction of $x$-axis. The dipole position is shown with the star in~\cref{fig:trans}(a). Injection of a short [full-width half-maximum of 7 electromagnetic wave periods at frequency $\omega a_0/(2\pi c)=0.417$] broadband pulse covering the frequency range $\omega a_0/(2\pi c)\in[0.417,0.4837]$ guaranteed excitation of all potentially guided states for any considered background refractive index combinations. 
	For accurate calculation of transmission, we used the simulation time equal to 1250 electromagnetic wave periods ensuring that all the energy coupled into the crystal is absorbed by PML domains.

	\section*{References}
	

	\pagebreak
	
	\section*{Acknowledgments}
	This work was supported by Army Research Office (ARO) grants: W911NF-16-1-0270 and W911NF-11-1-0297.
	
	\section*{Author contributions}
	All authors contributed significantly to the work presented in this paper.
	
	\section*{Competing financial interests}
	The authors declare no competing financial interests.

\newpage

\onecolumngrid

\setcounter{figure}{0}
\renewcommand{\thefigure}{S\arabic{figure}}
\addtocounter{equation}{-1}
\renewcommand{\theequation}{S.\arabic{equation}}

\section*{Supplementary materials for Reconfigurable Topological Photonic Crystal}

\subsection*{Topological properties of the structure}
\label{sec:topo}

\begin{figure}[!b]
	\includegraphics[width = 0.6\columnwidth, clip = true, trim = {0 0 0 0}]{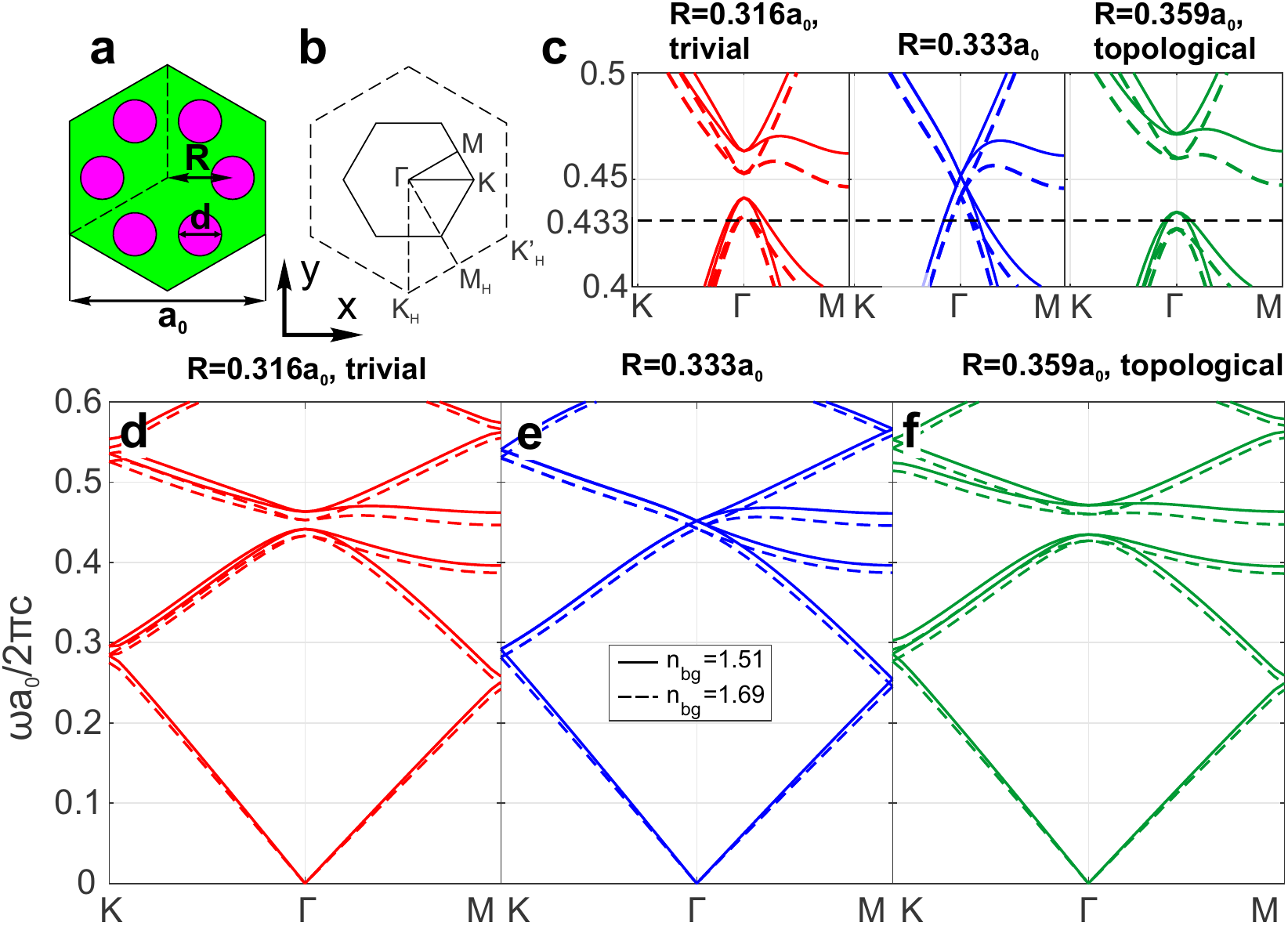}
	\caption{(a)~Schematic illustration of the unit cells of the triangular lattice (shaded in green) and the honeycomb lattice (dashed lines) with the geometrical parameters specified in the text. (b)~The first Brillouin zones of the triangular lattice (solid lines) and the honeycomb lattice (dashed lines). The high symmetry points of the triangular lattice are $\Gamma = [k_x^{(\Gamma)}, k_y^{(\Gamma)}] = [0,   0]$, $K=[4\pi/(3a_0),0]$, and $M = [\pi/a_0,\pi/(\sqrt{3}a_0)]$. High symmetry points of the honeycomb lattice are, $K_H=[0,-4\pi/(\sqrt{3}a_0)]$, $M_H = [\pi/a_0,-\sqrt{3}\pi/a_0]$, and $K'_H = [2\pi/a_0,-2\pi/(\sqrt{3}a_0)]$.  (f)--(h) Full photonic band structures, and zooms on the band-gap regions~(c)--(e) of the honeycomb lattice (triangular lattice with $R=a_0/3$---blue) exhibiting symmetry protected double Dirac cones~(d),~(g); triangular lattice with $R=0.316a_0$ (red) and opened trivial gap~(c),~(f); and topological triangular lattice with $R=0.359a_0$ (green) with non-trivial gap~(e),~(h). Solid lines and dashed lines correspond to the background refractive indices of 1.51 and 1.69, respectively.}
	\label{fig:bands}
\end{figure}

Let us consider a 2D photonic crystal (PC) with a structure of a honeycomb lattice of infinitely long silicon pillars with diameter $d$ (refractive index of silicon is taken to be 3.5) surrounded by a uniform material with the refractive index $n_{\textrm{bg}}$ schematically presented in~\cref{fig:fig1}(a) in the main text of the manuscript. The lattice constant of this honeycomb lattice is equal to $a$ and the unit cell containing two pillars is schematically shown in~\cref{fig:bands}(a) (dashed lines). In our 2D model, the PC structure is invariant along the $z$-axis, and we only analyze the behavior of the transverse-magnetic (TM) modes because, simple geometrical modifications of the crystal lattice allow for introduction of topological properties for this light polarization~\cite{wu15}. It is important to notice that the honeycomb lattice described above is equivalent to a larger triangular lattice containing six silicon pillars in a single unit cell. The dimensions of the unit cell of the triangular lattice are: the lattice constant $a_0=\sqrt{3}a$, the spacing between the unit cell center and the center of the pillar $R=a_0/3$, and the pillar diameter used here is $d=2/9a_0$. Description of the system in terms of a larger triangular lattice is required to characterize the topologically trivial and non-trivial PC geometries.

Band diagrams for the TM modes of the structure discussed above were calculated using the plane-wave-expansion (PWE) method for background refractive indices of 1.51 and 1.69, and are presented in Figs.~\ref{fig:bands}(d) and~(g). The allowed eigen-frequencies $\omega$ of the structure are shown in function of the wavevectors lying along the high-symmetry directions in the first Brillouin zone~(BZ) of the large triangular structure shown in~\cref{fig:bands}(b). The Floquet periodicity used for PC analysis assumes the phase difference $\phi = \vec{k} a_n$ on the opposite unit cell boundaries, where $a_n$ denotes one of the lattice vectors. For the periodicity along the $x$-direction, this phase can be expressed as $\phi = k_x a_0 = (k_x^{(1^{st}\textrm{BZ})} + \frac{2\pi}{a_0} m) a_0$. Here, $k_x$ denotes any of the allowed $k$-vectors, $k_x^{(1^{st}\textrm{BZ})}$ is its equivalent from the first BZ, and $m\in[0,1,2,\dots]$ denotes the number of full field oscillations in the unit cell. We observe the presence of a doubly-degenerate Dirac cone at the $\Gamma$ point, which is a result of folding of the bands from the larger BZ corresponding to the smaller unit cell of the honeycomb lattice [shown by dashed lines in \cref{fig:bands}(b)]~\cite{barik16}. The Dirac cones are located at points $K_H$ and $K_H'$ of the honeycomb lattice BZ. The doubly-degenerate Dirac cones are the consequence of the $C_6$ symmetry of the honeycomb lattice~\cite{wu15}.

Now, we modify the geometry of the PC by varying the pillar spacing $R$. This transformation does not affect the $C_6$ system symmetry but can lead to changes in the topological properties of the PC band structure. For values of $R\neq a_0/3$, the lattice can no longer be described as a honeycomb lattice, and therefore, introduction of the triangular lattice with a larger unit cell was necessary. Variations of $R$  lead to a spectral gap opening at the $\Gamma$ point, where the Dirac cones were located. Although the band-gap is opened regardless of whether we increase or decrease the pillar spacing, the topological character of the gap is different in both cases. Smaller values of $R$ result in weaker coupling between the adjacent meta-molecules and the topology of the system remains trivial. On the contrary, increased values of $R$ result in stronger inter-cell coupling and the gap has a non-trivial topological character (see next section of the Supplementary materials). The band diagrams for the topologically trivial system with $R=0.316a_0$, and for the non-trivial configuration with $R=0.359a_0$ are shown in Figs.~\ref{fig:bands}(c),~(f) and~\ref{fig:bands}(e),~(h), respectively.

The conducting topological edge states exist at an interface between the materials with overlapping band-gaps but different topological orders, defined by the topological invariant known as Chern number~\cite{hatsugai93}. The number of edge states, residing inside the bulk band-gap, is determined by the difference between Chern numbers of the two materials. The Chern number of the crystal lattice studied here can be evaluated using Hamiltonians derived from $k \cdot P$ theory~\cite{wu15,xu16} or tight-binding approximation~\cite{barik16}. In the following section, we present the tight-binding approach to the Chern number calculation.

\subsection*{Tight-binding model}

The photonic crystal considered in the main text of the manuscript is composed of a unit cell made of six silicon pillars, referred later as a meta-molecule (MM). For the TM-modes of the system, the $z$-component of the electric field $\psi$ in the unit cell can be represented as a combination of Wannier states:
\begin{equation}
\psi = \sum_{X=A,B...F} A_X \psi_X e^{i\vec{k}\cdot\vec{r}},
\end{equation}
where $\psi_X$ is the field distribution localized at the atom $X$, $A_X$ is the corresponding complex amplitude, $\vec{k} = [k_x, k_y]$ is the light wavevector, and $\vec{r}=(x,y)$ denotes the position.

\begin{figure}[!b]
	\includegraphics[width = 0.4\textwidth, clip = true, trim = {70 100 70 100}]{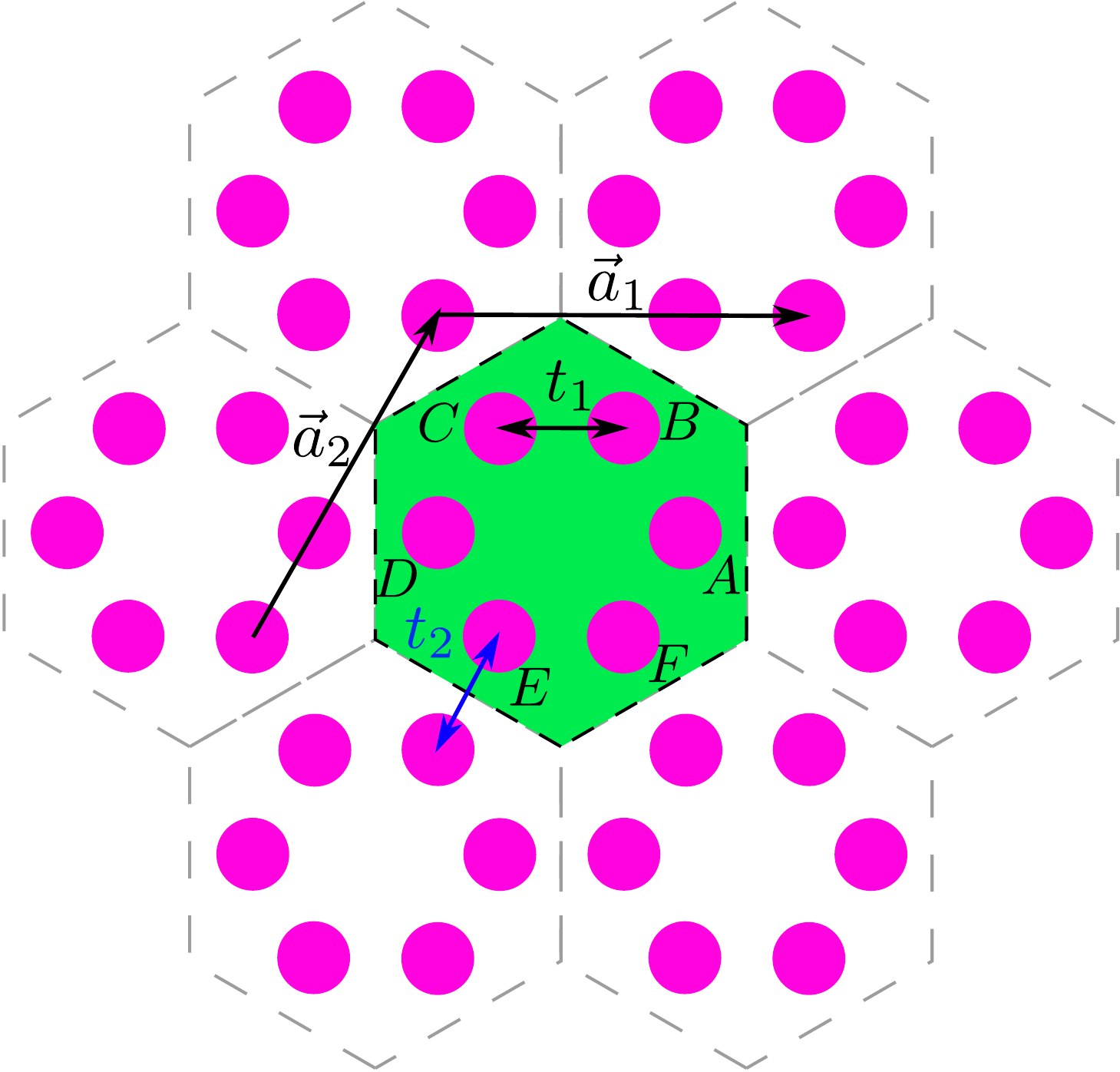}
	\caption{Meta-molecules (MMs) made of six silicon pillars forming the triangular lattice of our system. A unit cell containing a single MM is shown in green. The lattice vectors are given by: $a_1 = [\sqrt{3} ; 0]a$ and $a_2 = [\sqrt{3}/2; 3/2]a$. The intra-MM and inter-MM coupling coefficients are represented by $t_1$ and $t_2$, respectively.}
	\centering
	\label{fig:thisismyfigs2}
\end{figure}

In the matrix notation, the electric field distribution in our system can be represented using a six-element column-vector. The elements of the vector represent the field amplitudes at the corresponding atoms: 
\begin{equation}
\psi \rightarrow|\psi\rangle=\begin{bmatrix}
A_A, A_B, A_C,  A_D, A_E, A_F
\end{bmatrix}^T. 
\end{equation}
In this notation, the Wannier basis consists of the following six vectors: 
\begin{align}
|A\rangle=\begin{bmatrix}
1
\\ 0
\\ 0
\\ 0
\\ 0
\\ 0 
\end{bmatrix},
|B\rangle=\begin{bmatrix}0
\\ 1
\\ 0
\\ 0
\\ 0
\\ 0	
\end{bmatrix}, 
|C\rangle=\begin{bmatrix}0
\\ 0
\\ 1
\\ 0
\\ 0
\\ 0 
\end{bmatrix},
|D\rangle=\begin{bmatrix}0
\\ 0
\\ 0
\\ 1
\\ 0
\\ 0 	
\end{bmatrix},
|E\rangle=\begin{bmatrix}0
\\ 0
\\ 0
\\ 0
\\ 1
\\ 0 	
\end{bmatrix},
|F\rangle=\begin{bmatrix}0
\\ 0
\\ 0
\\ 0
\\ 0
\\ 1
\end{bmatrix},
\end{align}
and the wavefunction is written as:
\begin{equation}
|\psi\rangle=\sum_{X=A,B...F}A_X |X\rangle.
\end{equation}

The interaction Hamiltonian of the system can be described using nearest-neighbor hopping potentials. In this model, the interaction Hamiltonian $H_W$ (written in the basis of Wannier states) can be separated into the intra-MM coupling term, $H_1$, and the inter-MM coupling term, $H_2$.
The coupling energy between the nearest-neighbor atoms in the same MM is denoted by $t_1$, and its contribution to the Hamiltonian is:
\begin{equation}
H_1=-t_1\begin{bmatrix}
0& 1 & 0 & 0 & 0 &1 \\ 
1& 0 &  1& 0 & 0 &0\\ 
0& 1 &  0& 1 & 0 &0 \\ 
0& 0 &  1&  0& 1 &0 \\ 
0& 0 &  0&  1& 0 &1 \\ 
1& 0 &  0&  0& 1 & 0
\end{bmatrix}
\end{equation}
The interaction energy between the neighboring atoms from different MMs is denoted by $t_2$, and its contribution to the Hamiltonian is described by:
\begin{equation}
H_2=-t_2\begin{bmatrix}
0& 0 & 0 & e^{i\vec{k}\cdot\vec{a}_1} & 0 &0 \\ 
0& 0 &  0& 0 & e^{i\vec{k}\cdot\vec{a}_2} &0\\ 
0& 0 &  0& 0 & 0 &e^{i\vec{k}\cdot(\vec{a}_2-\vec{a}_1)} \\ 
e^{-i\vec{k}\cdot\vec{a}_1}& 0 &  0&  0& 0 &0 \\ 
0& e^{-i\vec{k}\cdot\vec{a}_2} &  0&  0& 0 &0 \\ 
0& 0 &  e^{-i\vec{k}\cdot(\vec{a}_2-\vec{a}_1)}&  0& 0 & 0
\end{bmatrix},
\end{equation}
where the $e^{i\vec{k}\vec{r}_{XY}}$ terms in the interaction between the $X$ and $Y$ atoms in neighboring MMs (the centers of the MMs are separated by the vector $\vec{r}_{XY}$) appear as a result of the Floquet periodic boundary conditions. 
The full interaction Hamiltonian $H_W$ = $H_1$+$H_2$ is given by:

\begin{equation}
H_W=-\begin{bmatrix}
0& t_1 & 0 & t_2e^{i\vec{k}\cdot\vec{a}_1} & 0 &t_1 \\ 
t_1& 0 &  t_1& 0 & t_2e^{i\vec{k}\cdot\vec{a}_2} &0\\ 
0& t_1 &  0& t_1 & 0 &t_2e^{i\vec{k}\cdot(\vec{a}_2-\vec{a}_1)} \\ 
t_2e^{-i\vec{k}\cdot\vec{a}_1}& 0 &  t_1&  0& t_1 &0 \\ 
0& t_2e^{-i\vec{k}\cdot\vec{a}_2} &  0&  t_1& 0 &t_1 \\ 
t_1& 0 &  t_2e^{-i\vec{k}\cdot(\vec{a}_2-\vec{a}_1)}&  0& t_1 & 0
\end{bmatrix}.
\end{equation}

Let us first analyze the system at the $\Gamma$ point ($\vec{k}=\vec{0}$). The eigen-energies are computed to be:
$E_s=-2t_1-t_2$, $E_p=-t_1+t_2$, $E_d=t_1-t_2$, and $E_f=2t_1+t_2$. They correspond to the eigen-states of the system:

\renewcommand{\arraystretch}{1.4}
\begin{align}
|s\rangle&=\begin{bmatrix}
\;\;\,\,1\;\;\,\,
\\ 1
\\ 1
\\ 1
\\ 1
\\ 1 
\end{bmatrix},&
|p_+\rangle&=\begin{bmatrix}1
\\ e^{\frac{i\pi}{3}}
\\ e^{\frac{2i\pi}{3}}
\\ e^{\frac{3i\pi}{3}}
\\ e^{\frac{4i\pi}{3}}
\\ e^{\frac{5i\pi}{3}}	
\end{bmatrix},& 
|d_+\rangle&=\begin{bmatrix}1
\\ e^{\frac{2i\pi}{3}}
\\ e^{\frac{4i\pi}{3}}
\\ e^{\frac{6i\pi}{3}}
\\ e^{\frac{8i\pi}{3}}
\\ e^{\frac{10i\pi}{3}}
\end{bmatrix},&
|p_-\rangle&=\begin{bmatrix}1
\\ e^{\frac{-i\pi}{3}}
\\ e^{\frac{-2i\pi}{3}}
\\ e^{\frac{-3i\pi}{3}}
\\ e^{\frac{-4i\pi}{3}}
\\ e^{\frac{-5i\pi}{3}}	
\end{bmatrix},&
|d_-\rangle&=\begin{bmatrix}1
\\ e^{\frac{-2i\pi}{3}}
\\ e^{\frac{-4i\pi}{3}}
\\ e^{\frac{-6i\pi}{3}}
\\ e^{\frac{-8i\pi}{3}}
\\ e^{\frac{-10i\pi}{3}}	
\end{bmatrix},&
\renewcommand{\arraystretch}{1.6}
|f\rangle&=\begin{bmatrix}\;\;\,\,1\;\;\,\,
\\-1
\\ 1
\\-1
\\ 1
\\-1
\end{bmatrix}.
\end{align}

\renewcommand{\arraystretch}{1}
The energy levels $E_p$ and $E_d$ are doubly degenerate. The eigen-states of the system at the $\Gamma$ point with the lowest and highest energies resemble the atomic orbitals $s$ and $f$, respectively. The eigen-states with intermediate energies, have a spin character and can be represented as a linear combination of atomic orbitals $p_x$, $p_y$, $d_{xy}$, and $d_{x^2-y^2}$: 
$p_\pm = \frac{1}{\sqrt{2}}( p_x \pm i p_y)$,
$d_\pm = \frac{1}{\sqrt{2}}( d_{x^2+y^2} \pm i d_{xy})$.
The field distributions for the eigen-states $|p\pm\rangle$ have the phase $e^{\pm i\phi}$, where the angle $\phi$ is measured in the coordinate frame centered in the middle of the unit cell. These states have a strong $p$ character. Similarly, field distributions for the eigen-states $|d\pm\rangle$ have the phase $e^{\pm 2 i\phi}$, and these states have a strong $d$ character. Both $p$ and $d$ states possess rotating (vortex-like) phase distributions analogous to the orbital angular momentum (pseudo-spin). The minus and plus sign corresponds to the clockwise (spin-up) and counterclockwise (spin-down) phase rotation direction, respectively.

In order to investigate the interaction between the spin-up and spin-down states, we express the Hamiltonian $H_W$ in the basis of $|s\rangle$, $|p_\pm\rangle$, $|d_\pm\rangle$, and $|f\rangle$ orbitals. The new Hamiltonian is obtained by a standard procedure of change of the basis and is given by:
\begin{equation}
H_{\textmd{spdf}} = M^{-1} H_W M,\nonumber
\end{equation} 
where $M$ is a $6\times6$ matrix of the form: 
\begin{equation}
M=\Big[
\;|s\rangle;\, |p_+\rangle;\,|d_+\rangle;\,|p_-\rangle;\,|d_-\rangle;\,|f\rangle
\;\Big],\nonumber
\end{equation}
whose columns represents the new basis vectors ($s$, $p_\pm$, $d_\pm$, $f$) in terms of the old basis vectors (Wannier functions). The Hamiltonian in the new basis has the form:
\small
\begin{align}
&H_{\textrm{spdf}}=\nonumber\\
&\left[
\begin{BMAT}{cccccc}{cccccc}
-2t_1-t_2+\frac{3}{4}a^2 t_2|\vec{k}|^2\;\;\;& \frac{\sqrt{3}}{2}at_2(ik_x-k_y) & 0 & \frac{\sqrt{3}}{2}at_2(ik_x+k_y)   & 0 &0 \\ 
-\frac{\sqrt{3}}{2}at_2(ik_x+k_y) & t_2-t_1 -\frac{3}{4}a^2 t_2|\vec{k}|^2\;\;\;& -\frac{\sqrt{3}}{2}at_2(ik_x-k_y) & 0 & 0 &0\\ 
0& \frac{\sqrt{3}}{2}at_2(ik_x+k_y)  & t_1-t_2+\frac{3}{4}a^2 t_2|\vec{k}|^2\;\;\;& 0 & 0 &\frac{\sqrt{3}}{2}at_2(ik_x-k_y)  \\ 
-\frac{\sqrt{3}}{2}at_2(ik_x-k_y)& 0 &  0&  t_2-t_1-\frac{3}{4}a^2 t_2|\vec{k}|^2\;\;\;& -\frac{\sqrt{3}}{2}at_2(ik_x+k_y) &0 \\ 
0& 0 &  0&  \frac{\sqrt{3}}{2}at_2(ik_x-k_y)& t_1-t_2+\frac{3}{4}a^2 t_2|\vec{k}|^2 \;\;\;&\frac{\sqrt{3}}{2}at_2(ik_x+k_y) \\ 
0& 0 &  -\frac{\sqrt{3}}{2}at_2(ik_x+k_y)&  0& -\frac{\sqrt{3}}{2}at_2(ik_x-k_y)& 2t_1+t_2-\frac{3}{4}a^2 t_2|\vec{k}|^2
\addpath{(1,1,1)rrrruuuulllldddd}
\end{BMAT}
\right]
\label{eqn:Hspdf}
\end{align}
\normalsize
where the terms $e^{i\vec{k}\vec{r}_{XY}}$ have been expressed using the Taylor expansion up to the terms linear (quadratic) in $k_x$ and $k_y$ in the off-diagonal (diagonal) elements.

Our main interest is to investigate the interaction between the spin states $|p_\pm\rangle$ and $|d_\pm\rangle$ because their interaction leads to the topological protection in the system. Therefore, in the following, we will consider a reduced $4\times4$ Hamiltonian containing only contributions from  $|p_\pm\rangle$ and $d_\pm\rangle$ states [highlighted in~\cref{eqn:Hspdf}]. This Hamiltonian has a block diagonal form~\cite{wu2}:
\begin{equation}
H_{\textrm{pd}}=\begin{bmatrix}
H_+& 0\\ 
0& H_-
\end{bmatrix},
\label{eqn:ham}
\end{equation}
where $H_\pm = \frac{\sqrt{3}}{2}t_2a(k_x\sigma_y \pm k_y\sigma_x)+(t_2-t_1-\frac{3}{4}a^2 t_2|\vec{k}|^2)\sigma_z$,
and $\sigma_x$, $\sigma_y$, and $\sigma_z$ are the Pauli matrices.

The Hamiltonian $H_{\textrm{pd}}$ has the same form as the Bernevig-Hughes-Zhang Hamiltonian describing the quantum spin-Hall effect in a quantum well systems~\cite{Bernevig1757}. Thus, our structure is capable of supporting optical pseudo-spin states.
The term $t_1-t_2$ is proportional to the energy difference at the $\Gamma$ point between the $|p_\pm\rangle$, $|d_\pm\rangle$ states. Depending on the sign of $t_1-t_2$, the character of the bands might be inverted leading to a nontrivial topological properties of the system. For $t_1>t_2$, the intra-MM coupling dominates and the energy of $|p_\pm\rangle$ bands is lower than the energy of $|d_\pm\rangle$ states, leading to the trivial state of the system. On the contrary, for $t_2>t_1$, the inter-MM coupling is dominant, leading to inversion to the $|p_\pm\rangle$ and $|d_\pm\rangle$ bands at the $\Gamma$ point and resulting in non-trivial topological properties~\cite{wu15}. When $t_2=t_1$, the $p$ and $d$ bands are degenerate at the $\Gamma$ point, resulting in the formation of the Dirac cone.

\begin{figure}[!t]
	\includegraphics[width = 0.65\textwidth, clip = true, trim = {0 0 0 0}]{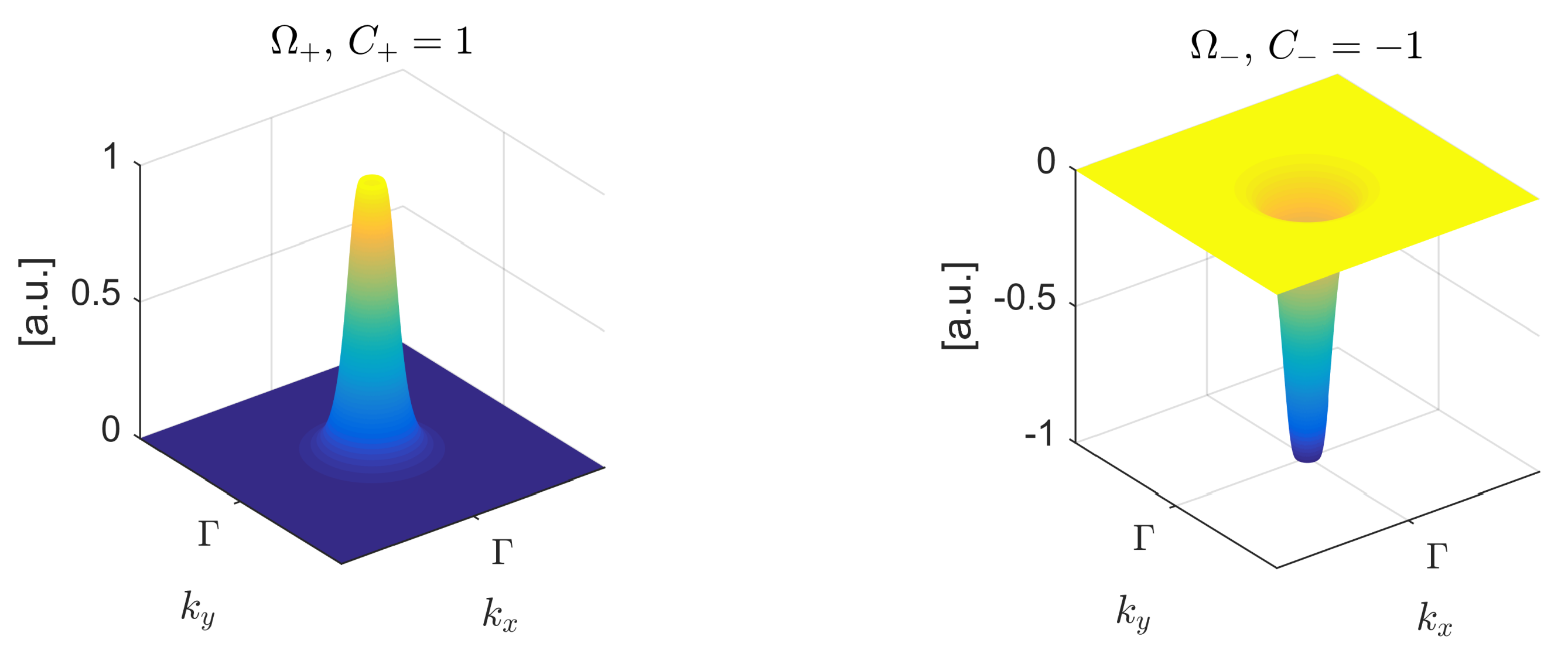}
	\caption{Berry curvatures $\Omega_\pm$ and Chern numbers $C_\pm$ of corresponding bands in the vicinity of the $\Gamma$ point for the system with $R=0.359a_0$ and $n_{\textrm{bg}}=1.51$.}
	\label{fig:thisismyfig1}
\end{figure}

Topological properties of any structure can be inferred from a topological invariant called Chern number. Chern number describes the global [computed in the entire first Brillouin zone~(BZ)] topological characteristics of the field distributions in a given energy band. In order to compute the Chern number, first we calculate the dispersion of light propagating in our topological structure. Using the PWE method, we obtain the $\omega(\vec{k})$ surfaces for the first 6 bands of the system together with the corresponding electric field distributions $E_z(\vec{r},\vec{k}) = \psi_k$. As mentioned above, we are interested only in states $|p_\pm\rangle$ and $|d_\pm\rangle$ showing the pseudo-spin character. For each of the energy bands, we compute the Berry connection $\vec{A}^{\pm} = [A_x^{\pm};A_y^{\pm}] = \iint_{\textrm{unit cell}} \psi_k^{\pm*} \nabla_k \psi_k^{\pm}\; \textrm{d}x\textrm{d}y $, where minus and plus signs correspond to the spin-up and spin-down states, respectively, and $\nabla_k$ denotes a two-dimensional gradient operator in the $k$-space. Based on the Berry connection, we can compute the Berry curvature as:
\begin{equation}
\Omega_\pm  = \frac{\partial A_y^{\pm}}{\partial k_x} - \frac{\partial A_x^{\pm}}{\partial k_y} , 
\end{equation}
Plots of the Berry curvature in the vicinity of the $\Gamma$ point for our topological structure are shown in \cref{fig:thisismyfig1}.
Finally, the Chern number is calculated as: 
\begin{equation}
C_{\pm} = \frac{1}{2\pi i} \iint_{1^{\textrm{st}}\textrm{BZ}} \Omega_\pm \textrm{d}k_x\textrm{d}k_y. 
\end{equation}

In our numerical approach, the eigen-spectrum of the Hamiltonian $H_{\textrm{pd}}$ given by~\cref{eqn:ham} is fitted to the $\omega(\vec{k})$ curves obtained from the PWE method allowing us to determine the $t_1$ and $t_2$ parameters. Then, the eigen-states of $H_{\textrm{pd}}$ are used to compute the Berry curvature and the Chern number using the approach presented in Ref.~\cite{200515018}.

\subsection*{Reconfigurability enabled by liquid crystals}

\begin{figure}[!t]
	\centering
	{\includegraphics[width = 0.198\columnwidth, clip = true, trim = {0 0 0 0}]{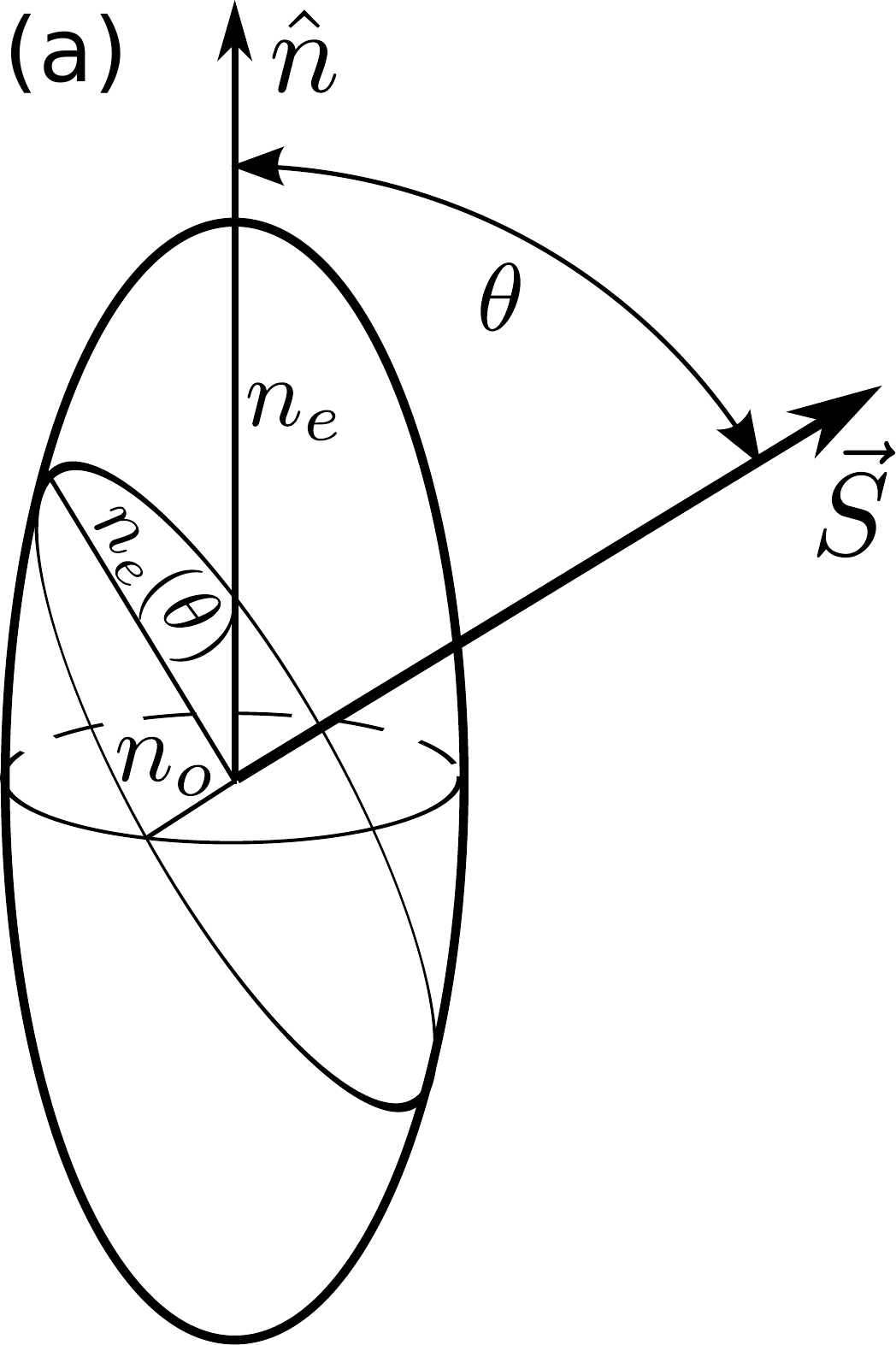}} \;\;\;
	{\includegraphics[width = 0.372\columnwidth, clip = true, trim = {0 0 0 0}]{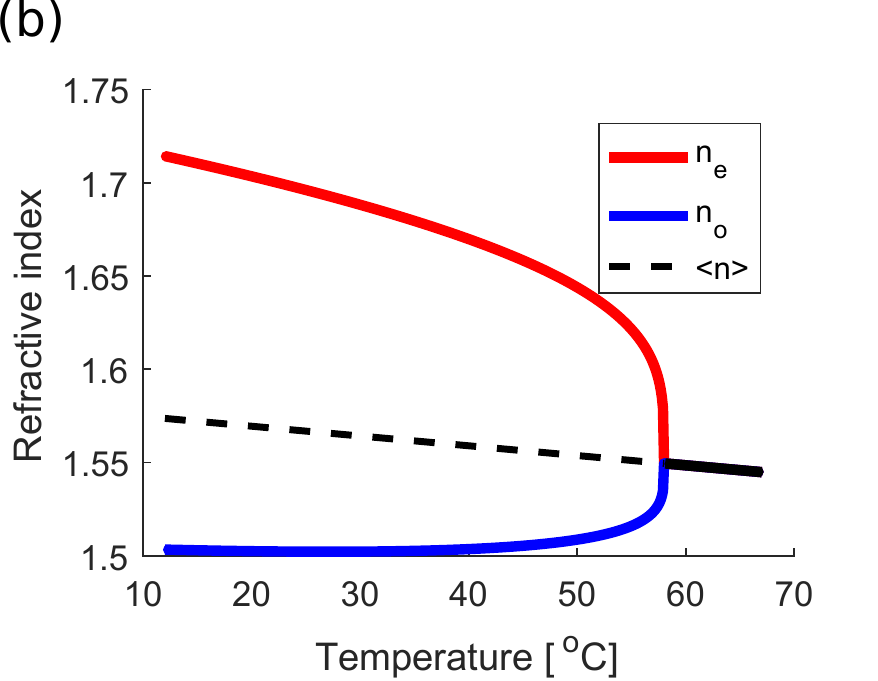}}
	\caption{(a)~Schematic illustration of a liquid crystal molecule showing the refractive indices along the long (director) axis $n_e$, along the short axis (ordinary index) $n_o$, and the extraordinary refractive index $n_e(\theta)$. The angle between the director axis $\hat{n}$ and the light propagation direction $\vec{S}$ is called $\theta$. (b) Dependence of the refractive indices $n_o$ and $n_e$ on temperature $T$ for the $E7$ liquid crystal at 1550~nm obtained using model from Ref.~\cite{li05b}.}
	\label{fig:LCTemp}
\end{figure}

Liquid crystals (LCs) are a state of matter that simultaneously exhibit properties of liquids (fluidity) and crystals (anisotropy resulting from orientation order of the molecules). LCs offer a possibility of refractive index tuning with unprecedented amplitude reaching 10\%~\cite{khoo93}. The elongated nematic LC molecules, schematically shown in~\cref{fig:LCTemp}(a), are preferentially aligned along a particular direction, called director, resulting in uniaxial anisotropy of the LC medium. The refractive index experienced by the extraordinarily polarized eigen-wave,  $n_e(\theta)$, changes depending on the angle $\theta$ between the director $\hat{n}$ and the light propagation direction $\vec{S}$:
\begin{equation}
n_e(\theta) = \frac{n_e n_o}{ \left[  n_e^2 \cos^2(\theta) + n_o^2 \sin^2(\theta) \right]^{\nicefrac{1}{2}}  }, 
\end{equation}
where $n_o$ denotes the ordinary refractive index and $n_e$ is the index of the wave polarized along the optical axis of the molecule. In this paper, we consider only the two extreme orientations of the LC molecules: (i) perpendicular to the PC pillars, for which the TM-polarized wave experiences the background index $n_{\textrm{bg}} = n_o$; and (ii) parallel to the PC pillars, for which $\theta = 90$\textdegree\ and $n_{\textrm{bg}} = n_e$, as explained below.

{\color{black}The PC structure studied in the main text of the manuscript can be bound from top and bottom by metal plates, as illustrated in~\cref{fig:fig1}(a) in the main text of the manuscript, that serve two goals. Firstly, for sufficiently small separation between the plates, the field distribution inside the PC structure remains uniform in the $z$-direction. Alternatively, the electrodes can be located at a distance from the PC slab. The space between electrodes and the slab can be filled with the LC which has much lower index than the PC itself. In this case, the light confinement stems from the total internal reflection between the high-index silicon slab and the low-index cladding. In both cases, these 3-dimensional~(3D) systems can be well approximated using 2D $z$-invariant modeling.}
This approximated treatment does not take into account the losses resulting from light penetrating into the metal regions and additional optimization is required to minimize these losses. Our 2D PC supports transverse-electric and transverse-magnetic modes. Here, we are only interested in the TM modes because the studied structure exhibits topological properties only for the this light polarization~\cite{wu15}. The TM mode has non-zero magnetic field components $H_x$ and $H_y$ lying in the $x$--$y$ plane perpendicular to the silicon pillars and an electric field component $E_z$ along the $z$-axis parallel to the pillars. 

Secondly, the {\color{black} conductive} plates serve as electrodes supplying an external electric field along the $z$-direction that allows for LC reorientation~\cite{khoo93,weirich10}. Due to the birefringent nature of LCs, the electric field exerts a torque on the LC molecules and, in case of positive birefringence ($\Delta n = n_e - n_o >0$), reorients them along the field direction. 
In our structure, the surface of the electrodes is prepared in such a way that the anchoring energy holds the surface molecules aligned parallel to the electrodes. In the absence of the electric field, the bulk molecules align along the surface molecules to minimize the global LC free energy, as shown in~\cref{fig:fig1}(d) in the main text of the manuscript. In this case, the TM-polarized wave is the ordinary wave and it experiences the background refractive index $n_{\textrm{bg}} = n_o$, irrespectively of the propagation direction in the $x$--$y$ plane and the in-plane director orientation. Application of an external electric field, with the amplitude above the threshold of the Fr\'{e}edericksz transition, results in rotation of the LC molecules~\cite{PhysRevA.40.6099}, as shown in~\cref{fig:fig1}(b) in the main text of the manuscript. Then, the director lies along the $z$-axis and is perpendicular to the light propagation direction ($\theta=90$\textdegree). In this configuration, the background refractive index experienced by the TM-polarized wave (extraordinary wave) is $n_{\textrm{bg}} = n_e$, irrespectively of the in-plane propagation direction. 

Orientations of the LC molecules at intermediate values of the angle $\theta$, in addition to being difficult to achieve experimentally, also break the reflection symmetry of the PC structure with respect to the $x$--$y$ plane. As a result, the TM-polarized wave excites both the ordinary and the extraordinary waves, and experiences polarization rotation due to the birefringence of the LC background. Systems with intermediate angles $\theta$ are not considered here and require further investigation.

Rotation of the LC molecules takes a finite amount of time due to their inertia and viscosity~\cite{doi:10.1080/15421406.2013.789428}. Typically, this switching time is of the order of microseconds allowing for device operation at MHz frequencies. Recently, sub-microsecond switching times were also reported~\cite{doi:10.1080/15421406.2011.569456,PhysRevLett.111.107802,Mohammadimasoudi:14}. 
In this paper, we use the $E7$ LC at the room temperature $T=25$\textdegree C which permits background index changes  in the range between $n_o=1.51$  and $n_e=1.69$~\cite{li05b,weirich10}.
Additional tunability can be added exploiting the fact that the values of $n_o$ and $n_e$ can be modified by temperature tuning, as seen in~\cref{fig:LCTemp}(b). For low temperatures, the LC birefringence is high, and the reorientation of the molecules leads to large changes in the background refractive index. At higher temperatures, the tunability is possible only in a limited range of background refractive indices, and therefore only a part of the parameter space shown in Figs.~\ref{fig:trans}(b)--(d) in the main text of the manuscript can be explored. Finally, for the temperatures above the nematic/isotropic phase transition point, no tunability is possible, as the LC becomes an isotropic liquid.


\end{document}